\documentclass[a4paper, 11pt]{article}

\usepackage{graphics,graphicx}
\usepackage{amssymb,amsmath}
\usepackage{color}
\usepackage{enumitem}

\newtheorem{thm}{Theorem}
\newtheorem{clm}{Claim}

\newtheorem{cor}{Corollary}
\newtheorem{lemma}{Lemma}

\def \no {\noindent}

\title{Chromatic bounds for some classes of $2K_2$-free graphs}

\author{T. Karthick\thanks{Corresponding author. Computer Science Unit, Indian Statistical
Institute, Chennai Centre, Chennai-600113, India.
E-mail: {karthick@isichennai.res.in}} \and Suchismita Mishra\thanks{Department of Mathematics, Indian Institute of Technology Madras, Chennai-600036, India.}}

\begin{document}
\maketitle

\begin{abstract}
A hereditary class
 $\mathcal{G}$ of graphs is $\chi$-{\it bounded} if there is a $\chi$-{\it binding function}, say
$f$ such that $\chi(G) \leq f(\omega(G))$, for every $G \in
\cal{G}$, where $\chi(G)$ ($\omega(G)$) denote the chromatic (clique)
number of $G$. It is known that for every $2K_2$-free graph $G$, $\chi(G) \leq \binom{\omega(G)+1}{2}$,
and the class of ($2K_2, 3K_1$)-free graphs does not admit a linear $\chi$-binding function.
In this paper,  we are interested in classes of $2K_2$-free graphs that admit a linear $\chi$-binding function. We show that the class of ($2K_2, H$)-free graphs, where $H\in \{K_1+P_4, K_1+C_4, \overline{P_2\cup P_3}, HVN,  K_5-e,  K_5\}$ admits a linear $\chi$-binding function. Also, we show that some superclasses of $2K_2$-free graphs are  $\chi$-bounded.
\end{abstract}

\no{\bf Keywords}. Chromatic number; clique number; graph classes; $2K_2$-free graphs.

%%%%%%%%%%%%%%%%%%%%%%%%%%%%%%%%%%%%%%%%%%%%%%%%%%%%%%%%%%%%%%%%%%%%%%%

\section{Introduction}

 All graphs in this paper are simple, finite and undirected.  For notation and terminology that are not defined here, we refer to
West \cite{West}.    Let $P_n$,
$C_n$, $K_n$ denote the induced path, induced cycle and complete graph
on $n$ vertices respectively. Let $K_{p,q}$ be the complete bipartite graph with classes of size $p$ and $q$. If $\cal{F}$ is a family of
graphs, a graph $G$ is said to be \emph{$\cal{F}$-free} if it contains no
induced subgraph isomorphic to any member of $\cal{F}$.  If $G_{1}$ and $G_{2}$ are two vertex disjoint graphs, then
  their {\it union} $G_1\cup G_2$ is the graph with $V(G_1\cup G_2)$
  $=V(G_1)\cup V(G_2)$ and $E(G_{1}\cup G_{2})$ $=$ $E(G_1)\cup
  E(G_2)$. Similarly, their {\it join} $G_{1}+G_{2}$ is the graph with
  $V(G_1+G_2)$ $=$ $V(G_1)\cup V(G_2)$ and $E(G_1 + G_2)$
  $=$ $E(G_1)\cup E(G_2)$$\cup \{(x,y)\mid x\in V(G_1), ~y\in
  V(G_2)\}$.    For any positive integer $k$, $kG$ denotes the union
  of $k$ graphs each isomorphic to $G$. For a graph $G$,  the complement of $G$ is denoted by $\overline{G}$.

A {\it proper coloring} (or simply {\it coloring}) of a graph $G$ is an assignment of colors to the vertices of $G$ such that no two adjacent
vertices receive the same color. The minimum number of colors required to
color $G$ is called the {\it chromatic number} of $G$, and is denoted by
$\chi(G)$. A {\it clique} in a graph $G$ is a set of vertices that are
pairwise adjacent in $G$. The {\it clique number}
of $G$, denoted by $\omega(G)$,  is the  size of a maximum clique in $G$.
Obviously, for any graph $G$, we have $\chi(G) \geq \omega(G)$.
The existence of triangle-free graphs with large chromatic number (see \cite{Mycielski} for a construction of such graphs) shows that for a general class of graphs, there is no upper bound on
the chromatic number as a function of clique number.

A graph $G$ is called {\it perfect} if $\chi(H)= \omega(H)$, for every
 induced subgraph $H$ of $G$; otherwise it is called {\it imperfect}.
A hereditary class
 $\mathcal{G}$ of graphs is said to be $\chi$-{\it bounded} \cite{Gyarfas} if there exists a function
$f$ (called a $\chi$-{\it binding function} of $\cal{G}$) such that $\chi(G) \leq f(\omega(G))$, for every $G \in
\cal{G}$.  If $\cal{G}$ is  the class of $H$-free graphs for some graph $H$,
 then $f$ is denoted by $f_H$.  We refer to \cite{Randerath-Schiermeyer-survey} for an extensive survey of
$\chi$-bounds for various classes of graphs.

The class of $2K_2$-free graphs and its related classes have been well studied in various contexts in the literature; see \cite{BLS1999}.
 Here, we would like to focus on showing $\chi$-binding functions for some classes of graphs related to $2K_2$-free graphs.
Wagon \cite{Wagon} showed that the class of $mK_2$-free graphs admits an $O(x^{2m-2})$ $\chi$-binding function for all
$m \geq 1$. In particular, he showed that $ f_{2K_2}(x) = \binom{x+1}{2}$, and the best known lower bound is $\frac{R(C_4 , K_{x+1})}{3}$,
where $R(C_4 , K_{x+1})$ denotes the smallest $k$ such that every graph on $k$
vertices contains either a clique of size $x + 1$ or the complement of the graph
contains a $C_4$ \cite{Gyarfas}. This lower bound is non-linear
because Chung \cite{chung} showed that $R(C_4 , K_t)$ is at least $t^{1 +\epsilon}$ for some $\epsilon > 0$.
It is interesting to note that Brause et al. \cite{BRSV} showed that the class of ($2K_2, 3K_1$)-free graphs does not admit a linear $\chi$-binding function. It follows that the class of ($2K_2, H$)-free graphs, where $H$ is any $2K_2$-free graph with independence number $\alpha(H) \geq 3$, does not  admit a linear $\chi$-binding function.

\begin{figure}
\centering
 \includegraphics{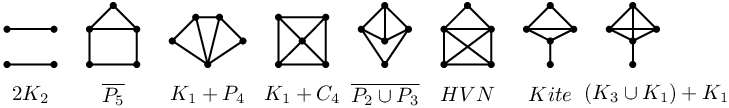}
\caption{Some special graphs.}\label{fig:1}
\end{figure}

Here we are interested in  classes of $2K_2$-free graphs that admit a linear $\chi$-binding function, in particular, some classes of $2K_2$-free graphs that admit a
`special' linear $\chi$-binding function $f(x) = x+c$, where $c$ is an integer, that is, $2K_2$-free graphs
${G}$ such that $\chi(G) \leq \omega(G) + c$.  If $c=1$, then this special upper
bound is called the {\it Vizing bound} for the chromatic number, and is well studied in the literature; see \cite{KM, Randerath-Schiermeyer-survey} and the references therein. Brause et al. \cite{BRSV} showed that  if $G$ is a connected ($2K_2, K_{1, 3}$)-free graph with independence number $\alpha(G) \geq 3$, then $G$ is perfect. It follows from a result of \cite{K} that if $G$ is a ($2K_2$, paw)-free graph, then either $G$ is perfect or $\chi(G)= 3$ and $\omega(G) =2$ (see also \cite{BRSV}). Nagy and Szentmikl\'ossy (see \cite{Gyarfas}) showed that if $G$ is a ($2K_2, K_4$)-free graph, then $\chi(G) \leq 4$.
 Blaszik et al. \cite{BHPT} and independently Gy\'arf\'as \cite{Gyarfas} showed that if $G$ is ($2K_2, C_4$)-free graph, then
$\chi(G) \leq \omega(G)+1$, and the equality holds if and only if $G$ is a split-graph. It follows from a result of \cite{KM} that if $G$ is a ($2K_2, K_4-e$)-free graph, then $\chi(G) \leq \omega(G)+1$. Fouquet et al. \cite{Fouquet} showed that if $G$ is a ($2K_2, \overline{P_5}$)-free graph, then $\chi(G) \leq \left\lfloor\frac{3\omega(G)}{2}
\right\rfloor$, and the bound is tight. Brause et al. \cite{BRSV} showed that if $G$ is a ($2K_2, K_1+P_4$)-free graph, then $\chi(G) \leq 2\omega(G)$.

\begin{table}
\centering

\begin{tabular}{|l|lll|}

\hline
{\small~~~~~~~Graph class $\cal{C}$} & \multicolumn{3}{c|}{{\small  $\chi$-bound for $G \in \cal{C}$}}\\
\hline

($2K_2, \overline{P_5}$)-free graphs & $\lfloor\frac{3\omega(G)}{2}
\rfloor$ & \cite{Fouquet} & \\

($2K_2, C_5$)-free graphs &  $\omega(G)^{3/2}$ &\cite{Hoang} &\\

($2K_2, K_1+P_4$)-free graphs & $\omega(G) +1$ & (Corollary~\ref{Bound-2K2-Gem}) & \\

($2K_2, K_1+C_4$)-free graphs & $\omega(G) +5$ & (Corollary~\ref{Bound-2K2-wheel}) & \\

($2K_2, \overline{P_2 \cup P_3}$)-free graphs & $\omega(G) +1$ & (Corollary~\ref{Bound-2K2-paraglider}) & \\

($2K_2$, $HVN$)-free graphs & $\omega(G)+3$ & (Corollary~\ref{Bound-2K2-HVN}) & \\

($2K_2, K_5-e$)-free graphs & $\omega(G)+4$ & (Corollary~\ref{Bound-2K2-K5-e}) &\\

($2K_2, K_5$)-free graphs & $2\omega(G)+1 \leq 9$ & (Corollary~\ref{Bound-2K2-K5}) & \\

($2K_2, X$)-free graphs & $\binom{\omega(G)+1}{2}$ & \cite{Wagon}  & \\

\hline
\end{tabular}
\caption{Known chromatic bounds for ($2K_2, H$)-free graphs, where $H$ is any
$2K_2$-free graph on $5$ vertices with $\alpha(H)= 2$, and the graph $X \in \{Kite, K_4\cup K_1, (K_3 \cup K_1)+K_1\}$.}
\label{tab:1}
\end{table}

In this paper, by using structural results, we show that the class of ($2K_2, H$)-free graphs, where $H\in \{K_1+P_4, K_1+C_4, \overline{P_2\cup P_3}, HVN, K_5-e\}$
 admits a special linear $\chi$-binding function $f(x) = x+c$, where $c$ is an integer; see Figure~\ref{fig:1}. We also show that the class of ($2K_2, K_5$)-free graphs admits a linear $\chi$-binding function.
Table~1 shows the known chromatic
bounds for a ($2K_2, H$)-free graph $G$, where $H$ is any
$2K_2$-free graph on $5$ vertices with $\alpha(H) =2$. Some of the cited bounds are
consequences of much stronger results available in the literature.
Finally, we show $\chi$-binding functions for some superclasses of $2K_2$-free graphs.

\section{Notation, terminology, and preliminaries}

Let $G$ be a graph, with vertex-set $V(G)$ and edge-set $E(G)$. For $x \in V(G)$, $N(x)$ denotes the set of all neighbors of $x$ in
$G$. For any two disjoint subsets $S,~T \subseteq V(G)$,
  $[S, T]$ denotes the set of edges
  $\{e \in E(G)\mid e \mbox{ has one end}$ in $S \mbox{ and the other
  in}~ T\}$. Also, for $S \subseteq V(G)$, let $G[S]$ denotes the subgraph induced by
  $S$ in $G$, and for convenience we simply write $[S]$ instead of $G[S]$. Note that  if $H_1$ and $H_2$ are any two graphs, and if $G$ is ($H_1, H_2$)-free, then $\overline{G}$ is ($\overline{H_1}, \overline{H_2}$)-free. For any integer $k$, we write $[k]$ to denote the set $\{1, 2, \ldots, k\}$.

A \emph{diamond} or a  $K_4-e$ is the graph with vertex set $\{a, b, c, d\}$ and edge set $\{ab, bc, cd, ad, bd\}$. A \emph{paw} is the graph  with vertex set $\{a, b, c, d\}$ and edge set $\{ab, bc, ac, ad\}$. See Figure \ref{fig:1} for some of the other special graphs used in this paper.

A graph $G$ is a {\it split graph} if its vertex set $V(G)$ can be partitioned into two sets $V_1$ and $V_2$ such that $V_1$ is a clique and $V_2$ is an independent set.   In \cite{FH}, F\"oldes and Hammer showed that a graph $G$ is a split graph if and only if $G$ is ($2K_2, C_4, C_5$)-free.
A graph $G$ is a {\it pseudo-split graph} \cite{MP} if $G$ is ($2K_2, C_4$)-free. The class of pseudo-split graphs generalizes the class of split graphs.

 A {\it $k$-clique covering} of a graph $G$ is a partition $(V_1, V_2, \ldots, V_k)$ of $V(G)$ such that $V_i$ is a clique, for each $i \in \{1, 2, \ldots, k\}$.  The {\it clique covering number} of the graph $G$, denoted by $\theta(G)$, is the minimum integer $k$ such that $G$ admits a $k$-clique covering. An {\it independent/stable}  set in a graph $G$ is a set of vertices that are
pairwise non-adjacent in $G$. The {\it independence number} of $G$, denoted by $\alpha(G)$,  is the  size of a maximum independent set in $G$.
Clearly, for any graph $G$, we have $\chi(G) = \theta(\overline{G})$ and $\omega(G) = \alpha(\overline{G})$.

Let $G$ be a graph on $n$ vertices $v_1, v_2, \ldots , v_n$, and let
$H_1, H_2, \ldots, H_n$ be any $n$ vertex disjoint graphs. Then an
{\it expansion} $G(H_1, H_2, \ldots, H_n)$  of $G$ \cite{CK-2} is the graph
obtained from $G$ by

~(i) replacing the vertex $v_i$ of $G$ by $H_i$, $i = 1, 2, \ldots ,
n$, and

(ii) joining the vertices $x \in H_i$, $y \in H_j$ iff $v_i$ and
$v_j$ are adjacent in $G$.

An expansion is also called a {\it composition}; see \cite{West}. If
$H_i$'s are complete, it is called a {\it complete expansion} of
$G$. By a result of Lov\'asz \cite{Lovasz},  if $G, H_1, H_2, \ldots, H_n$ are perfect, then $G(H_1, H_2, \ldots, H_n)$ is perfect.

We also use the following known results:
\begin{enumerate}
\itemsep=-.2em
\item[(R1)]  {\bf Seinsche (\cite{Seinsche})}: {\it If $G_1$ and $G_2$ are $P_4$-free, then $G_1 \cup G_2$ and $G_1+G_2$ are $P_4$-free.}

\item[(R2)] {\bf Seinsche (\cite{Seinsche})}: {\it Every $P_4$-free graph is perfect.}

\item[(R3)] {\bf Chudnovsky et al. (\cite{Chudnovsky})} ({\textsc{The Strong Perfect Graph Theorem}} {\bf (SPGT)}):  \emph{A graph is perfect if and only if it
contains no odd hole (chordless cycle) of length at least $5$ and no
odd antihole (complement graph of a hole) of length at least $5$.}

\item[(R4)] {\bf Choudum et al. (\cite{CKS})}: \emph{Let $\cal{G}$ and $\cal{F}$ be hereditary classes of graphs where
$\cal{F}$ admits  a linear $\chi$-binding function. If there exists a
constant $k$ such that for any $G\in \cal{G}$, $V(G)$ can be
partitioned into $k$ subsets $V_{1}, V_{2}, \ldots, V_{k}$, where
$[V_i]\in \cal{F}$ for each $i \in \{1, \ldots, k\}$, then $\cal{G}$ has a
linear $\chi$-binding function.}

\item[(R5)]{\bf Blazsik et al. (\cite{BHPT})}: \emph{For every pseudo-split graph $G$, $\chi(G) \leq \omega(G) + 1$.}

 \item[(R6)]{\bf Brause et al.  (\cite{BRSV})}: \emph{For every ($2K_2$, paw)-free graph $G$, $\chi(G) \leq \omega(G) + 1$.}

 \item[(R7)]{\bf Karthick and Maffray (\cite{KM})}: \emph{For every ($2K_2$, diamond)-free graph $G$, $\chi(G) \leq \omega(G) + 1$.}

\end{enumerate}

\section{Linearly $\chi$-bounded  $2K_2$-free graphs}
In this section, we show that the class of ($2K_2, H$)-free graphs, where $H\in \{K_1+P_4, K_1+C_4, \overline{P_2\cup P_3}, K_5-e, HVN, K_5\}$
is linearly $\chi$-bounded. Note that the class of ($2K_2, K_1+C_4$)-free graphs  and the class of ($2K_2, \overline{P_2\cup P_3}$)-free graphs generalize the class of ($2K_2, C_4$)-free graphs or pseudo-split graphs. Also the class of ($2K_2, K_5-e$)-free graphs, the class of ($2K_2, \overline{P_2\cup P_3}$)-free graphs, and the class of ($2K_2, HVN$)-free graphs generalize the class of ($2K_2, K_4-e$)-free graphs.

\subsection{The class of ($2K_2, K_1+P_4$)-free graphs}

First we prove a structure theorem for the complement graph  of a ($2K_2, K_1+P_4$)-free graph.

\begin{thm}\label{2k2-gem-comp-struc}
Let $G$ be an imperfect $(P_4 \cup K_1, C_4)$-free graph. Then $G$ is connected and there
exists a partition $(V_1, V_2)$ of $V(G)$ such that
 $V_1$ induces a perfect subgraph of $G$, and $V_2$ is a  clique.
\end{thm}

\no{\it Proof.} %If $G$ is perfect, then $(V_1:= V(G),
%V_2:= \emptyset, V_3:= \emptyset)$ is a required partition. If $G$ is imperfect, then
Let $G$ be an imperfect $(P_4 \cup K_1, C_4)$-free graph.
Since $G$ is ($P_4 \cup K_1$)-free, $G$ contains no hole of length at least $7$, and
 since $G$ is $C_4$-free, $G$ contain no anti-hole of length at least $7$. Thus, it
follows from SPGT \cite{Chudnovsky} that $G$ contains a $5$-hole (hole
of length~$5$), say $C$ with vertex-set
$\{v_1, v_2, v_3, v_4, v_5\}$, and  edge-set $\{v_1v_2, v_2v_3, v_3v_4, v_4v_5, v_5v_1\}$. Throughout this proof, we take all the subscripts of $v_i$ to be modulo 5.

\setcounter{clm}{0}
\begin{clm}\label{cl:1}
Any vertex $x \in V(G) \setminus V(C)$ is adjacent to at least two vertices in $C$.
\end{clm}
\no{\it Proof of Claim \ref{cl:1}.} Suppose not. If $x$ is not adjacent to any of the vertices in $C$, or if $x$ is adjacent to exactly one vertex in $C$, say to $v_1$,  then $\{v_2, v_3, v_4, v_5, x\}$ induces a $P_4 \cup K_1$ in $G$, which is a contradiction. $\Diamond$

\medskip
By Claim~\ref{cl:1}, $G$ is connected.

\begin{clm}\label{cl:2}
If $x \in V(G) \setminus V(C)$, then $[N(x) \cap V(C)]$ is isomorphic to a member of $\{K_2, P_3, C_5\}$.
\end{clm}
\no{\it Proof of Claim \ref{cl:2}.} Suppose not. Then by Claim~\ref{cl:1}, $N(x) \cap V(C)$ is either $\{v_{i}, v_{i+3}\}$ or $\{v_{i}, v_{i+1}, v_{i+3}\}$ or $\{v_{i}, v_{i+1}, v_{i+2}, v_{i+3}\}$, for some $i$. But then in all the cases, $\{v_i, x, v_{i+3}, v_{i+4}\}$ induces a $C_4$ in $G$, which is a contradiction. $\Diamond$

\medskip
For $i \in [5]$, let:
\begin{eqnarray*}
A_{i} &=& \{x\in V(G)\setminus V(C) \mid  N(x)\cap V(C) = \{v_i, v_{i+1}\}\}, \\
B_i &=& \{x\in V(G)\setminus V(C) \mid  N(x)\cap V(C) =\{v_{i-1}, v_i, v_{i+1}\}\}, \\
D &=& \{x\in V(G)\setminus V(C) \mid  N(x)\cap V(C) = V(C)\}.
\end{eqnarray*}
Moreover, let $A=A_1\cup\cdots\cup A_5$,  and  $B=B_1\cup\cdots\cup B_5$.
  Then by Claims \ref{cl:1} and \ref{cl:2}, we have $V(G) = V(C)\cup A\cup B \cup D$.

\begin{clm}\label{cl:Ai}
For each $i \in [5]$ ($i$ mod 5), the following hold:
\begin{itemize}
     \itemsep=-.3em
\item[{\rm(i)}] $A_i$ induces a $P_4$-free subgraph of $G$.
\item[{\rm(ii)}] $[A_i, A_{i+1}]$ is complete.
\item[{\rm(iii)}] If $A_i \neq \emptyset$ and $A_{i+1} \neq \emptyset$, then $A_i$ and $A_{i+1}$ are cliques in $G$.
\item[{\rm(iv)}] If $A_i \neq \emptyset$, then $A_{i+2} = \emptyset = A_{i-2}$.

\end{itemize}
\end{clm}
\no{\it Proof of Claim \ref{cl:Ai}.} (i) Suppose to the contrary that $[A_i]$ contains an induced $P_4$, say $P$. Then by the definition of $A_i$, $V(P) \cup \{v_{i+3}\}$ induces a $P_4 \cup K_1$ in $G$, a contradiction. So (i) holds.

Suppose that (ii) does not hold. Then there exist vertices $x \in A_i$ and $y \in A_{i+1}$ such that $xy \notin E(G)$. But, then $\{v_{i+3}, v_{i+4}, v_{i}, x, y\}$ induces a $P_4 \cup K_1$ in $G$, a contradiction.  So
(ii) holds.

Suppose that (iii) does not hold. Then up to symmetry, there exist two non-adjacent vertices $a$ and $b$ in $A_i$, and let $x \in A_{i+1}$.
Then by (ii), $ax, bx \in E(G)$.  But then  $\{a, b, x, v_i\}$ induces a $C_4$ in $G$, a contradiction.  So
(iii) holds.

Suppose that (iv) does not hold. Then there exist vertices $x \in A_i$ and $y \in A_{i+2} \cup A_{i-2}$. By symmetry, we may assume that $y \in A_{i+2}$. But then  $\{x, v_{i+1}, v_{i+2}, y\}$ induces a $C_4$ in $G$, if $xy \in E(G)$, and $\{x, v_{i+1}, v_{i+2}, y, v_{i+4}\}$ induces a $P_4 \cup K_1$ in $G$, if $xy \notin E(G)$, a contradiction. So (iv) holds. $\Diamond$

\begin{clm}\label{cl:Bi}
For each $i \in [5]$ ($i$ mod 5), the following hold:
\begin{itemize}
     \itemsep=-.3em
\item[{\rm(i)}] $\{v_i\}\cup B_i \cup D$ is a clique.
\item[{\rm(ii)}] $[B_i, B_{i+2}] = \emptyset = [B_i, B_{i-2}]$.
\item [{\rm(iii)}] $[B_i \cup B_{i+1} \cup B_{i+2} \cup B_{i+3}]$ is a perfect subgraph of $G$.
\end{itemize}
\end{clm}
\no{\it Proof of Claim \ref{cl:Bi}.} Suppose that (i) does not hold. Then there exist vertices $x, y \in B_i \cup D$ such that $xy \notin E(G)$.
But, then $\{x, v_{i-1}, y, v_{i+1}\}$ induces a  $C_4$ in $G$, a contradiction.
So (i) holds.

Suppose that (ii) does not hold. Then there exist vertices $x \in B_i$ and $y \in B_{i+2}\cup B_{i-2}$ such that $xy \in E(G)$. By symmetry, we may assume that $y \in B_{i+2}$. But, then $\{x, v_{i+4}, v_{i+3}, y\}$ induces a $C_4$ in $G$, a contradiction.  So
(ii) holds. $\Diamond$

It is clear that (iii) follows from (i), (ii), and by SPGT \cite{Chudnovsky}.

\begin{clm}\label{cl:AiBi}
For each $i \in [5]$ ($i$ mod 5), the following hold:
\begin{itemize}
     \itemsep=-.3em
\item[{\rm(i)}] $[A_i, B_i\cup B_{i+1}]$ are complete.
\item[{\rm(ii)}] $[A_i, B_{i+3}] = \emptyset$.
\item[{\rm(iii)}] If $x \in B_{i+2}\cup B_{i-1}$, then either $[\{x\}, A_i]$ is complete or $[\{x\}, A_i]= \emptyset$.
%\item[{\rm(iv)}] If $A_i$ is not a clique, then $[A_i,  B_{i+2}\cup B_{i-1}] = \emptyset$.
\end{itemize}
\end{clm}
\no{\it Proof of Claim \ref{cl:AiBi}.} Suppose that (i) does not hold. Then there exist vertices $x \in A_i$ and $y \in B_{i}\cup B_{i+1}$ such that $xy \notin E(G)$.  But, then $\{v_{i+2}, v_{i+3}, v_{i+4}, y, x\}$ induces a $P_4 \cup K_1$ in $G$, a contradiction.  So (i) holds.

Suppose that (ii) does not hold. Then there exist vertices $x \in A_i$ and $y \in B_{i+3}$ such that $xy \in E(G)$. But, then $\{x, v_{i}, v_{i+4}, y\}$ induces a $C_4$ in $G$, a contradiction.  So (ii) holds.

By symmetry, we may assume that $x \in B_{i+2}$. Suppose that (iii) does not hold.   Then there exist vertices $a$ and $b$ in $A_i$ such that $ax \in E(G)$ and $bx \notin E(G)$. Then since $\{v_{i+4}, v_{i+3}, a, x, b\}$ does not induce a $P_4 \cup K_1$, we have $ab \in E(G)$. But, then $\{v_{i+2}, a, x, b, v_{i+4}\}$ induces a $P_4 \cup K_1$, a contradiction. So (iii) holds.
$\Diamond$

\medskip
By Claim~\ref{cl:Ai}(iv), we may assume that $A \setminus (A_1\cup A_2) = \emptyset$.
%Now, by using the above claims, we prove the theorem in two  cases (the other cases are symmetric).
If $A_1 \neq \emptyset$ and $A_2 \neq \emptyset$ or if $A_1 \neq \emptyset$ is a clique and $A_2 = \emptyset$ or if $A_1 \cup A_2 = \emptyset$, then we define  $V_1:= \{v_1, v_3, v_4, v_5\} \cup B_1 \cup B_3\cup B_4 \cup B_5$ and $V_2:= \{v_2\} \cup A_1 \cup A_2 \cup B_2$. Then by the definitions of $B_i$ and by Claim~4(iii), $V_1$ induces a perfect subgraph of $G$. Also, by Claim~3(iii) and by Claim~5(i), $V_2$ is a clique in $G$. So $(V_1, V_2)$ is a required partition of $G$ and the theorem holds.

So, suppose that $A_1$ is not a clique. Let $a$ and $b$ be two vertices in $A_1$ that are non-adjacent. First, note that by Claim~\ref{cl:AiBi}(i), $[A_1, B_1 \cup B_2]$ is complete. Moreover:

\begin{clm}\label{cl:AiBi-rev}
We have the following:
\begin{itemize}
     \itemsep=-.3em
\item[{\rm(i)}]   $[A_1,  B_5] = \emptyset$.
\item[{\rm(ii)}]    $[B_1, B_2]$, $[B_1, B_5]$ and $[B_3, B_4]$ are complete.
\end{itemize}
\end{clm}
\no{\it Proof of Claim \ref{cl:AiBi-rev}.}
$(i)$: Suppose that (i) does not hold. Then there exists a vertex $x$ in $B_{5}$ and a vertex in $A_1$ that are adjacent. Then by Claim~\ref{cl:AiBi}(iii), $[\{x\}, A_1]$ is complete. In particular, $ax, ay \in E(G)$. But, then $\{x, a, b, v_{2}\}$ induces a $C_4$ in $G$, a contradiction. So (i) holds.

\noindent{$(ii)$}: If $[B_1, B_2]$ is not complete, then there exist vertices $x \in B_1$ and $y \in B_2$ such that $xy \notin E(G)$. But then $\{x, y, a, b\}$ induces a $C_4$ in $G$, a contradiction. So, $[B_1, B_2]$ is  complete.

If $[B_1, B_5]$ is not complete, then there exist vertices $x \in B_1$ and $y \in B_5$ such that $xy \notin E(G)$. Then since
$\{y, v_5, x, a, v_3\}$ or $\{y, v_5, x, b, v_3\}$ do not induce a $P_4 \cup K_1$, we have $ya, yb \in E(G)$. But then $\{y, a, b, v_2\}$ induces a $C_4$ in $G$, a contradiction. So, $[B_1, B_5]$ is  complete.

If $[B_3, B_4]$ is not complete, then there exist vertices $x \in B_3$ and $y \in B_4$ such that $xy \notin E(G)$.  Then by Claim~\ref{cl:AiBi}(ii), $ya, yb \notin E(G)$. Then since $\{v_5, y, v_3, x, a\}$ or $\{v_5, y, v_3, x, b\}$ do not induce a $P_4 \cup K_1$, we have $xa, xb \in E(G)$. But then $\{x, a, b, v_1\}$ induces a $C_4$ in $G$, a contradiction. So, $[B_3, B_4]$ is  complete. $\Diamond$

Now, we define  $V_1:= \{v_1, v_2, v_5\} \cup A_1 \cup B_1 \cup B_2 \cup B_5$ and $V_2:= \{v_3, v_4\} \cup B_3\cup B_4$.
Then by above claims, we see that $V_1$ induces a perfect subgraph of $G$ as it is a join of two perfect subgraphs induced by $\{v_1\} \cup B_1$ and $\{v_2, v_5\} \cup A_1 \cup B_2 \cup B_5$, and $V_2$ is a clique.  Hence the theorem is proved. \hfill{$\Box$}

\medskip
The following corollary is an improvement over that in \cite{BRSV}, where it is shown that for every
($2K_2, K_1+P_4$)-free graph $G$, $\chi(G) \leq 2\omega(G)$.

\begin{cor}\label{Bound-2K2-Gem}
Let $G$ be a  ($2K_2, K_1+P_4$)-free graph. Then $\chi(G) \leq \omega(G) +1$.
\end{cor}
\no{\it Proof}. Consider the complement $H$ of $G$. Then $H$ is a ($P_4 \cup K_1, C_4$)-free graph.

If $H$ is perfect, then $\theta(H) = \alpha(H)$, and the corollary holds.

If $H$ is imperfect, then by Theorem~\ref{2k2-gem-comp-struc}, $H$ is connected and there exists a partition ($V_1, V_2$) of  $H$ such that $V_1$ induces a perfect subgraph of $H$, and $V_2$ is a clique in $H$. So, $\theta(H) \leq \theta([V_1])+\theta([V_2]) =  \alpha([V_1]) + 1 \leq \alpha(H)+ 1$, and the corollary follows. \hfill{$\Box$}

The bound in Corollary~\ref{Bound-2K2-Gem} is tight. For example, consider the graph $G$ isomorphic to $C_5[K_t^c, K_t^c, K_t^c, K_t^c, K_t^c]$. Then $G$ is $(2K_2, K_1+P_4)$-free with $\omega(G) = 2$ and $\chi(G) = 3$.

\subsection{The class of ($2K_2, K_1+C_4$)-free graphs}

First we prove a structure theorem for the class of ($2K_2, K_1+C_4$)-free graphs.

\begin{thm}\label{2k2-4-wheel-struc}
Let $G$ be a connected ($2K_2, K_1+C_4$)-free graph. Then $G$ is either a pseudo-split graph or there
exists a partition $(V_1, \ldots, V_6)$ of $V(G)$ such that
\begin{enumerate}
\item[(i)] $[V_1]$ is either a pseudo-split graph of $G$ with $\omega([V_1]) \leq \omega(G) -1$ or  the complement of a bipartite graph of $G$, and
\item[(ii)] $V_i$ is an independent set,  for each $i \in \{2, \ldots, 6\}$. Moreover, if $V_1$ induces a pseudo-split graph of $G$, then $V_5 =\emptyset= V_6$.
\end{enumerate}
\end{thm}

\no{\it Proof.} Let $G$ be a connected ($2K_2, K_1+C_4$)-free graph.

If $G$ is $C_4$-free, then $G$ is a pseudo-split graph, and the theorem holds.

Suppose that $G$ contains an induced $C_4$, say $C$ with vertex-set
$L_0 := \{v_1, v_2, v_3, v_4\}$, and  edge-set $\{v_1v_2, v_2v_3, v_3v_4, v_4v_1\}$. Define sets $L_1:= \{y
\in V(G)\setminus L_0  \mid  y\mbox{ has a neighbor in } L_0\}$ and $L_2 :=
V(G) \setminus (L_0\cup L_1)$. Throughout this proof, we take all the subscripts of $v_i$ to be modulo 4.

\setcounter{clm}{0}
\begin{clm}\label{cl:L1}
If $x \in L_1$, then $|N(x) \cap L_0| \in \{1, 2, 3\}$.
\end{clm}
\no{\it Proof of Claim \ref{cl:L1}.} Otherwise, $L_0 \cup \{x\}$ induces a $K_1+C_4$ in $G$, a contradiction. $\Diamond$

\medskip
So, for any $x \in L_1$, there exists an index $j \in [4]$  such that $xv_j \in E(G)$ and $xv_{j+1} \notin E(G)$.
For $i \in [4]$, let:
\begin{eqnarray*}
W_{i} &=& \{x\in L_1 \mid  N(x)\cap L_0 = \{v_i\}\}, \\
X_i &=& \{x\in L_1 \mid  N(x)\cap L_0 =\{v_i, v_{i+1}\}\}, \\
Y_1 & = & \{x\in L_1 \mid  N(x)\cap L_0 = \{v_{1}, v_3\}\},\\
Y_2 & = & \{x\in L_1 \mid  N(x)\cap L_0 = \{v_{2}, v_4\}\},\\
Z_i &=& \{x\in L_1 \mid  N(x)\cap L_0 =\{v_{i-1}, v_i, v_{i+1}\}\}.
\end{eqnarray*}
Moreover, let $W=W_1\cup\cdots\cup W_4$,  $X=X_1\cup\cdots\cup X_4$, and $Z=Z_1\cup\cdots\cup Z_4$.
  Then, by Claim~\ref{cl:L1}, $V(G) = L_0 \cup W\cup X \cup Y_1\cup Y_2 \cup Z \cup L_2$. Now:

\begin{clm}\label{cl:L2}
The following hold:
\begin{itemize}
     \itemsep=-.3em
\item[{\rm(i)}] If $x \in W\cup X$, then $N(x) \cap L_2 = \emptyset$.
\item[{\rm(ii)}] $L_2$ is an independent set.
\end{itemize}
\end{clm}
\no{\it Proof of Claim \ref{cl:L2}.} (i) We may assume that $x \in W_1\cup X_1$, and suppose to the contrary that $y \in N(x) \cap L_2$.
Then $\{y, x, v_3, v_4\}$ induces a $2K_2$ in $G$, a contradiction.  So (i) holds.

Suppose that (ii) does not hold. Then there exist two adjacent vertices, say $x$ and $y$ in $L_2$. But, then $\{x, y, v_1, v_2\}$ induces a $2K_2$ in $G$, a contradiction.  So (ii) holds. $\Diamond$

\begin{clm}\label{cl:L1-Y}
For each $i \in [4]$ ($i$ mod 4), the following hold:
\begin{itemize}
     \itemsep=-.3em
\item[{\rm(i)}] $W_i\cup W_{i+1} \cup X_i$ is an independent set.
%\item[{\rm(ii)}] $[W_i, W_{i+1}] = \emptyset$.
\item[{\rm(ii)}] If $X_i \neq \emptyset$, then either $X_{i+1} = \emptyset$ or $X_{i+2} = \emptyset$.
\item[{\rm(iii)}] If $Z_i \neq \emptyset$, then $Z_{i+2} = \emptyset$.

\end{itemize}
\end{clm}
\no{\it Proof of Claim \ref{cl:L1-Y}.} We prove the claim for $i =1$.

(i) Suppose to the contrary that there exist two adjacent vertices, say $x$ and $y$ in $W_1\cup W_2 \cup X_1$.
Then $\{x, y, v_3, v_4\}$ induces a $2K_2$ in $G$, a contradiction. So (i) holds.

%Suppose that (ii) does not hold. Then there exist vertices $x \in W_1$ and $y \in W_{2}$ such that $xy \in E(G)$. But, then $\{x, y, v_3, v_4\}$ induces a $2K_2$ in $G$, a contradiction.  So (ii) holds.

Suppose that (ii) does not hold. Then there exist vertices $x_1 \in X_1$, $x_2 \in X_2$ and $x_3 \in X_3$. Then since $\{x_1, v_1, x_2, v_3\}$ or $\{x_1, v_1, x_3, v_3\}$ or $\{x_2, v_2, x_3, v_4\}$ do not induce a $2K_2$ in $G$, $\{x_1, x_2, x_3\}$ induces a triangle in $G$. But, then $\{x_1, v_2, v_3, x_3, x_2\}$ induces a $K_1+C_4$ in $G$, a contradiction. So (ii) holds.

Suppose that (iii) does not hold.  Then there exist vertices $x \in Z_1$ and $y \in Z_3$. But then  $\{v_1, v_2, y, v_4, x\}$ induces a $K_1+C_4$ in $G$, if $xy \in E(G)$, or $\{v_1, x, y, v_3\}$ induces a $2K_2$ in $G$, if $xy \notin E(G)$, a contradiction. So (iii) holds. $\Diamond$

\begin{clm}\label{cl:Yi}
For each $i\in \{1, 2\}$, $Y_i$ is a union of a clique and an independent set.
\end{clm}
\no{\it Proof of Claim \ref{cl:Yi}.} We prove the claim for $i =1$.  First, we show that $[Y_1]$ is $P_3$-free.
Suppose to the contrary that $[Y_1]$ contains an induced $P_3$, say $P$. Then by the definition of $Y_1$, $V(P) \cup \{v_1, v_3\}$ induces a $K_1+C_4$ in $G$, a contradiction. So, $[Y_1]$ is $P_3$-free, and hence it is a union of cliques.  Then since $G$ is $2K_2$-free, it follows that $Y_1$ is a union of a clique and an independent set, and the claim holds.  $\Diamond$

\medskip
By Claim~\ref{cl:Yi}, for each $i\in \{1, 2\}$, we define $Y_i := Y_i' \cup Y_i''$, where $Y_i'$ is a clique, and $Y_i''$ is an independent set.

\begin{clm}\label{cl:YiZi-1}
For each $i \in \{1, 2\}$,  $[Z_i\cup Z_{i+2}, Y_{3-i}]= \emptyset$.
\end{clm}
\no{\it Proof of Claim \ref{cl:YiZi-1}.} We prove the claim for $i =1$. Suppose to the contrary that there exist vertices, say $z \in Z_1 \cup Z_3$ and $y \in Y_2$ such that $zy \in E(G)$. But, then $\{v_1, v_2, y, v_4, z\}$ or $\{v_2, v_3, v_4, y, z\}$ induces a $K_1+C_4$ in $G$, a contradiction. So the claim holds. $\Diamond$

\begin{clm}\label{cl:YiZi-2}
For each $i \in \{1, 2\}$, if $Z_i\cup Z_{i+2} \neq \emptyset$, then $Y_{3-i}$ is an independent set.
\end{clm}
\no{\it Proof of Claim \ref{cl:YiZi-2}.} We prove the claim for $i =1$. Let $z \in Z_1\cup Z_3$.  Up to symmetry, we may assume that $z \in Z_1$. By Claim~\ref{cl:YiZi-1}, $[\{z\}, Y_2] = \emptyset$. Now, we show that $Y_2$ is an independent set. Suppose to the contrary that there exist adjacent vertices, say $p$ and $q$ in $Y_2$. Then since $[\{z\}, Y_2] = \emptyset$, we have $zp \notin E(G)$ and $zq \notin E(G)$. But, then $\{z, v_1, p, q\}$ induces a $2K_2$ in $G$, a contradiction.  So the claim holds. $\Diamond$

Now, by using Claim~\ref{cl:L1-Y}(iii), we prove the theorem in two cases.

\medskip
\no{\bf Case~1.} {\it Suppose that $Z_i = \emptyset$, for every $i \in [4]$.}

By Claim~\ref{cl:L1-Y}(ii) and by symmetry, we may assume that either $X_2\cup X_4 = \emptyset$ or $X_3 \cup X_4 = \emptyset$.
Then we define $V_1:= Y_1' \cup Y_2' \cup \{v_1, v_2\}$, $V_2:= Y_1'' \cup \{v_4\}$, $V_3:= Y_2'' \cup \{v_3\}$,
$V_4:= W_1 \cup W_2 \cup X_1 \cup L_2$. Further: If $X_2 \cup X_4 = \emptyset$, then we define $V_5:= W_3 \cup W_4 \cup X_3$ and $V_6 := \emptyset$; and if   $X_3 \cup X_4 = \emptyset$, then we define $V_5:= W_3 \cup X_2$ and $V_6 := W_4$.

Now, by Claims~\ref{cl:L2} and \ref{cl:L1-Y}(i), and by the definition of $Y_i'$'s and $Y_i''$'s, we see that $[V_1]$ is isomorphic to the complement of a bipartite graph, and $V_i$'s are independent sets, for each $i \in \{2, \ldots, 6\}$. So, $(V_1, \ldots, V_6)$ is a required partition of $V(G)$.

\medskip
\no{\bf Case~2.} {\it Suppose that $Z_i \cup Z_{i+1} \neq \emptyset$, for exactly one $i\in [4]$.}

We may assume up to symmetry that $i =1$ and $Z_1 \neq \emptyset$. Then by Claim~\ref{cl:YiZi-2}, $Y_2$ is an independent set. Then, we define $V_1:= N(v_1)$, $V_2:= W_2 \cup X_2 \cup L_2$, $V_3:= W_3 \cup W_4 \cup X_3$, $V_4:= Y_2 \cup \{v_1, v_3\}$, $V_5 := \emptyset$ and $V_6:= \emptyset$.
Then since $G$ is ($K_1+C_4$)-free, $V_1$ induces a pseudo-split graph in $G$. Also, $\omega([V_1]) \leq \omega(G) -1$. So, by Claims~\ref{cl:L2} and \ref{cl:L1-Y}(i), we see that $V_i$'s are independent sets, for each $i \in \{2, 3, 4\}$, and hence $(V_1, \ldots, V_6)$ is a required partition of $V(G)$.

This completes the proof of the theorem.
\hfill{$\Box$}

\begin{figure}[t!]
\centering
 \includegraphics{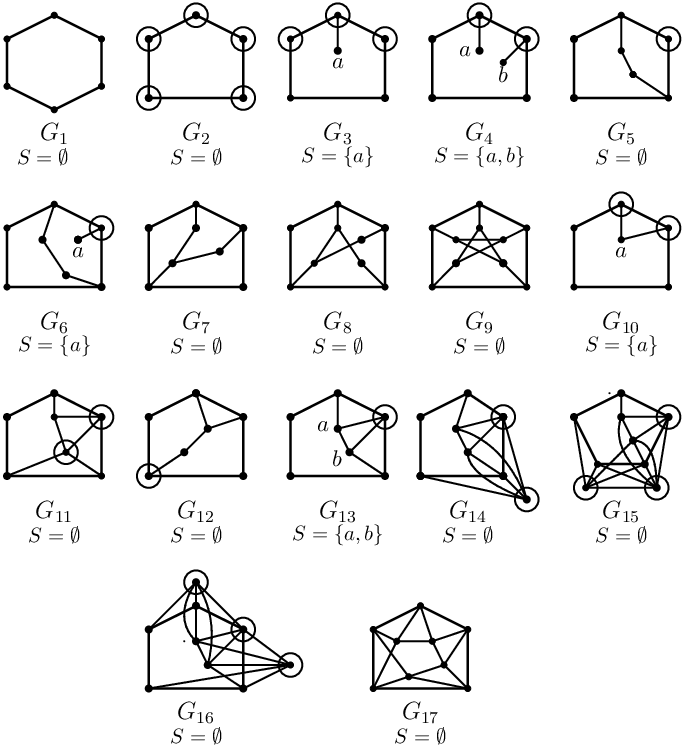}
\caption{Basic graphs used in Theorem \ref{structure
thm}.}\label{Basic-graphs-P2P3-C4}
\end{figure}

\begin{cor} \label{Bound-2K2-wheel}
Let $G$ be a  ($2K_2, K_1+C_4$)-free graph. Then $\chi(G) \leq \omega(G) +5$.
\end{cor}
\no{\it Proof}. Let $G$ be a  ($2K_2, K_1+C_4$)-free graph. We may assume that $G$ is connected.  We use Theorem~\ref{2k2-4-wheel-struc}.
If $G$ is a  pseudo-split graph, then,  by (R5), $\chi(G) \leq \omega(G) +1$.
So, suppose that $V(G)$ admits a partition as in Theorem~\ref{2k2-4-wheel-struc}.
 Now:

  (a) Suppose that $V_1$ induces a pseudo-split graph with $\omega([V_1]) \leq \omega(G) -1$. Then $V_5 = \emptyset = V_6$. So, $\chi(G) \leq \chi([V_1]) + 3$. Then by (R4) and (R5), $\chi(G) \leq \omega([V_1]) +1+ 3 \leq \omega(G)+3$.

(b) Suppose that $[V_1]$ is isomorphic to the complement of a bipartite graph, and $V_i$'s are independent sets, for each $i \in \{2, \ldots, 6\}$.
Then since $[V_1]$ is perfect, it follows  by (R4) that $\chi(G) \leq \omega([V_1]) + 5\leq \omega(G)+5$. Hence the corollary is proved. \hfill{$\Box$}

\subsection{The class of ($2K_2, \overline{P_2\cup P_3}$)-free graphs}

We use the following structure theorem for $(P_2 \cup P_3, C_4)$-free graphs proved in \cite{CK-2}.

\begin{thm}[\cite{CK-2}]\label{structure thm}
If $G$ is a connected $(P_2 \cup P_3, C_4)$-free graph, then $G$
is either chordal or there exists a partition $(V_1, V_2, V_3)$ of
$V(G)$ such that (1) $[V_1] \cong K_m^c$, for some $m \geq 0$, (2)
$[V_2] \cong K_t$, for some $t \geq 0$, (3) $[V_3]$ is isomorphic to
a graph obtained from one of the basic graphs $G_t$ ($1 \leq t \leq
17$) shown in Figure \ref{Basic-graphs-P2P3-C4} by expanding each
vertex indicated in circle by a complete graph (of order $\geq 1$),
(4) $[V_1, V_3] =\emptyset$, and (5) $[V_2, V_3 \setminus S]$ is
complete (see Figure \ref{Basic-graphs-P2P3-C4} for the set $S$).
\end{thm}

For $ t \in [17]$, let ${\cal{G}}_t$
denote the class of graphs obtained from $G_t$ (see Figure~\ref{Basic-graphs-P2P3-C4}) by the operations
stated in Theorem~\ref{structure thm}.

\begin{cor}\label{Bound-2K2-paraglider}
Let $G$ be a  ($2K_2, \overline{P_2\cup P_3}$)-free graph. Then $\chi(G) \leq \omega(G) +1$.
\end{cor}
\no{\it Proof}. Consider the complement $H$ of $G$. Then $H$ is a $(P_2 \cup P_3, C_4)$-free graph.

If $H$ is chordal, then $H$ is perfect and so $\theta(H) = \alpha(H)$, and the corollary holds.

Suppose that $H$ is not chordal. Let $H_1, H_2, \ldots, H_k$ ($k\geq 1$) denote the components of $H$. Then since $H$ is not chordal and since $H$ is $(P_2 \cup P_3, C_4)$-free, by Theorem~\ref{structure thm}, we may assume that there exists a component, say $H_1$ of $G$ such that $V(H_1)$ admits a partition $(V_1, V_2, V_3)$ as in Theorem~\ref{structure thm} where $[V_3]$ contains either a $C_5$ or a $C_6$, and  $[V_3] \in {\cal{G}}_t$, for $t \in [17]$. Then since $H$ is ($P_2\cup P_3$)-free, $H_i \cong K_1$, for each $i \in \{2, 3, \ldots, k\}$. So $\alpha(H) = \alpha(G_t)+|V_1|+(k-1)$. Now, $\theta(H) \leq \theta([V_1]) + \theta([V_2\cup V_3]) + (k-1) = |V_1| + \theta([V_2\cup V_3]) + (k-1) = \theta([V_2\cup V_3]) + \alpha(H)-\alpha(G_t)$. It is easily verified that $ \theta([V_2\cup V_3]) \leq \alpha(G_t)+1$. Hence, $\theta(H) \leq \alpha(H)+1$, and the corollary is proved. \hfill{$\Box$}

\medskip
 The graphs $C_5(\overline{K}_{n_1}, \overline{K}_{n_2}, \overline{K}_{n_3}, \overline{K}_{n_4}, \overline{K}_{n_5})$ show
that the bound in Corollary~\ref{Bound-2K2-paraglider} is tight.

\subsection{The class of ($2K_2,H$)-free graphs, $H \in  \{HVN, K_5-e\}$}

In order to prove our next results, we need the following notation.
Let $G$ be a connected graph that contains an induced diamond, say $D$, with vertex set $L_0:= \{v_1, v_2, v_3, v_4\}$ and
edge set $\{v_1v_2, v_2v_3, v_3v_4, v_4v_1, v_2v_4\}$. Define sets $L_1:= \{y
\in V(G)\setminus L_0  \mid  y\mbox{ has a neighbor in } L_0\}$ and $L_2 :=
V(G) \setminus (L_0\cup L_1)$. Moreover, let:
\begin{eqnarray*}
X_{i} &=& \{x \in L_1 \mid  N(x)\cap L_0 = \{v_i\}\};  i \in [3], \\
Y_1 &=& \{x \in L_1 \mid  N(x)\cap L_0 =\{v_1, v_{2}\}\},\\
Y_2 &=& \{x \in L_1 \mid  N(x)\cap L_0 =\{v_2, v_{3}\}\},\\
Z_1 & = & \{x\in L_1 \mid  N(x)\cap L_0 = \{v_{1}, v_3\}\},\\
Z_2 & = & \{x\in L_1 \mid  N(x)\cap L_0 = \{v_{1}, v_2, v_3\}\}.
\end{eqnarray*}

Then we have the following lemma, and we leave its proof as it can be routinely verified.

\begin{lemma}\label{2K2-struc-diamond}
  Let $G$ be a connected $2K_2$-free graph that contains an induced diamond $D$. Let $L_0$, subsets of $L_1$, and $L_2$ be defined as above.
Then the following hold:
\begin{enumerate}
\item[(1)] $V(G) = N(v_4) \cup \{v_4\} \cup X_1 \cup X_2 \cup X_3 \cup Y_1 \cup Y_2 \cup Z_1 \cup Z_2 \cup L_2$.
\item[(2)] We have either $X_1 = \emptyset$ or $X_3 = \emptyset$.
\item[(3)] $X_1 \cup X_2 \cup Y_1$, $Y_2$, $Z_1$ and $L_2$ are independent sets.
\item[(4)] $[X_1 \cup X_2 \cup X_3 \cup Y_1 \cup Y_2 \cup Z_1, L_2] = \emptyset$.
 \hfill{$\Box$}
\end{enumerate}
\end{lemma}

\begin{thm}\label{2k2-HVN-struc}
Let $G$ be a connected ($2K_2, HVN$)-free graph. Then $G$ is either a ($2K_2$, diamond)-free graph or there
exists a partition $(V_1, \ldots, V_4)$ of $V(G)$ such that
\begin{enumerate}
\item[(i)] $V_1$ induces a ($2K_2$, paw)-free graph of $G$ with $\omega([V_1]) \leq \omega(G) -1$, and
\item[(ii)] $V_i$ is an independent set,  for each $i \in \{2, 3, 4\}$.
\end{enumerate}
\end{thm}

\no{\it Proof.} Let $G$ be a connected ($2K_2, HVN$)-free graph.
If $G$ is diamond-free, then the theorem holds.
Suppose that $G$ contains an induced diamond, say $D$. We use Lemma~\ref{2K2-struc-diamond}. By (2) and by symmetry, we may assume that $X_3 = \emptyset$. Now, since $G$ is $HVN$-free, we have the following:

\begin{itemize}
  \item For any $v \in V(G)$, $N(v)$ induces a paw-free graph with $\omega([N(v)]) \leq \omega(G) -1$.
  \item $Y_2 \cup Z_2$ is an independent set (by using (3)).
\end{itemize}

 \no{}Define $V_1:= N(v_4)$, $V_2:= X_1 \cup X_2 \cup Y_1$, $V_3:= Y_2 \cup Z_2$, and $V_4:= Z_1 \cup L_2 \cup \{v_4\}$. Then by (3) and (4), and by the above properties, we see that $(V_1, \ldots, V_4)$ is a required partition of $V(G)$, and the theorem is proved.  \hfill{$\Box$}

 \begin{cor} \label{Bound-2K2-HVN}
   Let $G$ be a  ($2K_2, HVN$)-free graph. Then $\chi(G) \leq \omega(G) +3$.
 \end{cor}

 \no{\it Proof.}  Let $G$ be a connected ($2K_2, HVN$)-free graph. We use Theorem~\ref{2k2-HVN-struc}.

 If $G$ is a ($2K_2$, diamond)-free graph, then by (R8),   $\chi(G) \leq \omega(G) +1$, and the corollary holds.
 Suppose that $V(G)$ admits a partition as in Theorem~\ref{2k2-HVN-struc}. So, $\chi(G) \leq \chi([V_1]) +3$.  Since $\chi([V_1]) \leq \omega([V_1]) +1$ (by (R7)), we have $\chi(G) \leq \omega([V_1]) +1 +3 \leq \omega(G) +3$, as desired. \hfill{$\Box$}

 \begin{thm}\label{2k2-K5-e-struc}
Let $G$ be a connected ($2K_2, K_5-e$)-free graph. Then $G$ is either a ($2K_2$, diamond)-free graph or there
exists a partition $(V_1, \ldots, V_5)$ of $V(G)$ such that
\begin{enumerate}
\item[(i)] $V_1$ induces a ($2K_2$, diamond)-free graph of $G$ with $\omega([V_1]) \leq \omega(G) -1$, and
\item[(ii)] $V_i$ is an independent set,  for each $i \in \{2, 3, 4, 5\}$.
\end{enumerate}
\end{thm}

\no{\it Proof.} Let $G$ be a connected ($2K_2, K_5-e$)-free graph.
If $G$ is diamond-free, then the theorem holds.
Suppose that $G$ contains an induced diamond, say $D$. We use Lemma~\ref{2K2-struc-diamond}. By (2) and by symmetry, we may assume that $X_3 = \emptyset$.  Now, since $G$ is ($K_5-e$)-free, we have the following:

\begin{itemize}
  \item For any $v \in V(G)$, $N(v)$ induces a diamond-free graph with $\omega([N(v)]) \leq \omega(G) -1$.
  \item $Z_2$ is an independent set.
\end{itemize}

\no{}Define $V_1:= N(v_4)$, $V_2:= X_1 \cup X_2 \cup Y_1$, $V_3:= Y_2$, $V_4:= Z_1 \cup L_2 \cup \{v_4\}$, and $V_5:= Z_2$. Then by (3) and (4), and by the above properties, we see that $(V_1, \ldots, V_5)$ is a required partition of $V(G)$, and the theorem is proved.  \hfill{$\Box$}

\begin{cor} \label{Bound-2K2-K5-e}
   Let $G$ be a  ($2K_2, K_5-e$)-free graph. Then $\chi(G) \leq \omega(G) +4$.
 \end{cor}

 \no{\it Proof.}  Let $G$ be a connected ($2K_2, K_5-e$)-free graph. We use Theorem~\ref{2k2-K5-e-struc}.

 If $G$ is a ($2K_2$, diamond)-free graph, then by (R8),   $\chi(G) \leq \omega(G) +1$, and the corollary holds.
 Suppose that $V(G)$ admits a partition as in Theorem~\ref{2k2-K5-e-struc}. So, $\chi(G) \leq \chi([V_1]) +4$.  Since $\chi([V_1]) \leq \omega([V_1]) +1$ (by (R7)), we have $\chi(G) \leq \omega([V_1]) +1 +4 \leq \omega(G) +4$, as desired. \hfill{$\Box$}

\subsection{The class of ($2K_2, K_1+ H$)-free graphs, for any graph $H$}
\begin{thm} \label{general-3}
Let $H$ be any graph. Suppose that for every $(2K_2, H)$-free graph $G'$, $\chi(G') \leq f(\omega(G'))$.  Then for every  ($2K_2, K_1+H)$-free graph $G$,   we have $\chi(G) \leq 2f(\omega(G)-1)+ 1$.
\end{thm}

\no{\it Proof}. Let $G$ be a ($2K_2, K_1+H)$-free graph. If $G$ is an edgeless graph, then the theorem is obvious. So we may assume that there exist adjacent vertices, say $v_1$ and $v_2$ in $V(G)$. For each $i\in \{1, 2\}$, let
$A_i: = \{x \in V(G)\setminus \{v_1, v_2\} \mid N(x) \cap \{v_1, v_2\} = \{v_i\}\}$. Also, let $B: = \{x \in V(G)\setminus \{v_1, v_2\} \mid N(x) \cap \{v_1, v_2\} = \{v_1, v_2\}\}$ and $C:= V(G) \setminus (\{v_1, v_2\} \cup A_1 \cup A_2\cup B)$. Then, we have the following:

\begin{enumerate}
\item[(i)] Since $G$ does not induce a $K_1+H$, we have: for any $v \in V(G)$, $N(v)$ induces a $H$-free graph.
So, for each $i \in \{1, 2\}$,  $[A_i \cup B]$ is a $H$-free graph with $\omega([A_i \cup B]) \leq \omega(G) -1$.
\item[(ii)]  Since $G$ does not induce a $2K_2$, we see that $C$ is an independent set.
\end{enumerate}

Now, $\chi(G) \leq \chi([N(v_1)]) + \chi([A_2]) + \chi([C\cup \{v_1\}]) = \chi([A_1 \cup B \cup \{v_2\}]) + \chi([A_2]) + \chi([C\cup \{v_1\}])$. Since for every ($2K_2, H$)-free graph $G'$, $\chi(G') \leq f(\omega(G'))$, and since $C \cup \{v_2\}$ is an independent set (by (ii)),  we have, by (R4),  $\chi(G) \leq f(\omega(G) -1) + f(\omega(G)-1) + 1 = 2f(\omega(G)-1)+1$ (by (i)), as desired.  \hfill{$\Box$}

\begin{cor}\label{2K2-K5-bound} \label{Bound-2K2-K5}
Let $G$ be a  ($2K_2, K_5$)-free graph. Then $\chi(G) \leq 2\omega(G) +1 \leq 9$.
\end{cor}

\no{\it Proof}. Since $G$ is a ($2K_2$, $K_1+K_4$)-free graph, and since for every ($2K_2, K_4$)-free graph $G'$, $\chi(G') \leq \omega(G') +1 \leq 4$ (see \cite{Gyarfas}), the corollary follows by Theorem~\ref{general-3}.  \hfill{$\Box$}

\section{Superclasses of $2K_2$-free graphs}

In this section, we show that some superclasses of $2K_2$-free graphs are $\chi$-bounded.

If $G$ is a graph and if  $e:= uv$ is an edge in $G$, then we simply write $A(e)$ to denote the set of all vertices in $G$ that are not adjacent to both $u$ and $v$ in $G$.
The proof of the following theorem is very similar to the proof of Wagon \cite{Wagon} for the class of $2K_2$-free graphs, and we give it here for completeness.

\begin{thm} \label{general}
Let $\cal{H}$ be a class of graphs and let $G$ be any graph. Suppose that $\cal{H}$ is  $\chi$-bounded with $\chi$-binding function $f$.  Suppose that for every edge $e$ in $G$, $[A(e)] \in \cal{H}$. Then $\chi(G) \leq \binom{\omega(G)}{2}\cdot f(\omega(G))+\omega(G)$. %
%If for every edge $e$ in $G$, $\overline{N(e)}$ induces a perfect graph, then $\chi(G) \leq \frac{\omega(G)^3-\omega(G)^2+2\omega(G)}{2}$.
\end{thm}

\no{\it Proof}. Let $\omega := \omega(G)$ and let $K$ be a complete subgraph of $G$ with
$|K| = \omega$, and $V(K) = \{v_1, v_2, \ldots, v_{\omega}\}$. Then every vertex in $x\in V(G) \setminus V(K)$ is not adjacent to at least one vertex in $K$. Otherwise, $\{x\} \cup V(K)$ induces a clique of  size larger than $\omega$ which is a contradiction. For each $i, j \in  [\omega]$, $i \neq j$, let
$A_{ij}:= A(e_{ij})$, where $e_{ij}$ is the edge $v_iv_j$,  and let $B_i:= \{x \in V(G) \setminus V(K) \mid [\{x\}, V(K) \setminus \{v_i\}]~ \mbox{is~complete}\}$. Moreover, let $A := \cup A_{ij}$ and $B:= \cup B_i$. Then $V(G) = V(K) \cup A \cup B$.

Now, for each $i, j \in  [\omega]$, $i \neq j$, we have:
 \begin{enumerate}
 \item[(i)] Since  for every edge $e$ in $G$, $[A(e)] \in \cal{H}$, we have $[A_{ij}] \in \cal{H}$.
 \item[(ii)] $B_i \cup \{v_i\}$ is an independent set. If not, then there exist adjacent vertices, say $x$ and $y$ in $B_i$. But, then $\{x,y\} \cup (V(K)\setminus \{v_i\})$ induces a clique of size $\omega+1$, a contradiction.
 \end{enumerate}

       So, $\chi(G) \leq \sum_{\{i, j\} \subseteq [\omega]} \chi([A(e_{ij})]) + \sum_{i =1}^{\omega}\chi([B_i\cup\{v_i\}])$. Then by (i) and (ii), and by (R4), $\chi(G) \leq \sum_{\{i, j\} \subseteq [\omega]} f(\omega([A(e_{ij})])) + \sum_{i =1}^{\omega}\chi([B_i\cup\{v_i\}])$. Then since $\omega([A(e_{ij})]) \leq \omega$ and since $B_i \cup \{v_i\}$ is an independent set, for each $i, j \in [\omega]$, $i \neq j$,  we have $\chi(G) \leq \binom{\omega}{2}\cdot f(\omega)+\omega$, and the theorem is proved. \hfill{$\Box$}
    %
%    Then by (R8), $\chi(G) \leq \sum_{\{i, j\} \subseteq [\omega]} f(\omega([A_{ij}])) + \sum_{i =1}^{\omega}\chi([B_i\cup\{v_i\}])$. Then since $f(\omega([A_{ij}])) \leq f(\omega)$ and since $B_i \cup \{v_i\}$ is an independent set, for each $i \in [\omega]$, we have $\chi(G) \leq \binom{\omega}{2}\cdot f(\omega)+\omega$, and the theorem is proved. \hfill{$\Box$}

\medskip
Then we immediately have the following.

\begin{cor}[\cite{Wagon}]
Let $G$ be a  $2K_2$-free graph. Then $\chi(G) \leq \binom{\omega(G) +1}{2}$.
\end{cor}

\no{\it Proof}. Since $G$ is  $2K_2$-free, for each $i, j \in  [\omega]$, $i \neq j$, $A(e_{ij})$ is an independent set in $G$. So, $\omega([A(e_{ij})]) \leq 1$, and hence the corollary follows from the proof of Theorem~\ref{general}. \hfill{$\Box$}

\begin{cor} \label{general-2}
Let $G$ be any graph. If for every edge $e$ in $G$, $A(e)$ induces a perfect graph, then $\chi(G) \leq \frac{\omega(G)^3-\omega(G)^2+2\omega(G)}{2}$. \hfill{$\Box$}
\end{cor}

\begin{cor}
Let $G$ be a  $(P_2 \cup P_4)$-free graph. Then $\chi(G) \leq  \frac{\omega(G)^3-\omega(G)^2+2\omega(G)}{2}$.
\end{cor}

\no{\it Proof}. Since every $P_4$-free is perfect (by (R2)), the corollary follows from Corollary~\ref{general-2}.  \hfill{$\Box$}

\medskip
\no{\bf Acknowledgement}. The first author sincerely thanks Prof.~Ingo Schiermeyer for the fruitful discussions.

\end{document}